\documentclass[preprint,12pt]{elsarticle}

\usepackage{amssymb}

\journal{New Astronomy}

\begin{document}

\begin{frontmatter}



\title{Possible identifications of newly observed magnetar quasi-periodic oscillations as crustal shear modes}


\author{Hajime Sotani$^1$}
\ead{hajime.sotani@nao.ac.jp}
\author{Kei Iida$^2$}
\author{Kazuhiro Oyamatsu$^3$}

\address{
$^1$Division of Theoretical Astronomy, National Astronomical Observatory of Japan, 2-21-1 Osawa, Mitaka, Tokyo 181-8588, Japan\\
$^2$Department of Natural Science, Kochi University, 2-5-1 Akebono-cho, Kochi 780-8520, Japan\\
$^3$Department of Human Informatics, Aichi Shukutoku University, 2-9 Katahira, Nagakute, Aichi 480-1197, Japan
}

\begin{abstract}
Quasi-periodic oscillations (QPOs) discovered in soft-gamma repeaters (SGRs) 
are expected to help us to study the properties of matter in
neutron stars.  In earlier investigations, we identified the QPOs 
of frequencies below $\sim100$ Hz observed in giant flares of SGR 1806$-$20 
and SGR 1900+14 as the crustal torsional oscillations.  For this purpose, we 
calculated the frequencies of the fundamental torsional oscillations with
various angular indices $\ell$, by changing the stellar mass and 
radius.  In this work, we try to explain the additional QPO frequencies 
recently reported by Huppenkothen et al. \cite{QPO1,QPO2} within the same framework as before
except that we newly take into account the effect of electron screening, 
which acts to decrease the frequencies by a small amount.   Those QPOs 
were discovered in two different SGRs, i.e., SGR 1806$-$20 and SGR 
J1550$-$5418.   Then, we find that the newly observed QPO frequency in
SGR 1806$-$20 can be still identified as one of the frequencies of the 
fundamental torsional oscillations, while those in SGR J1550$-$5418 can 
also be explained in terms of the torsional oscillations although
the relevant angular indices are difficult to identify.
\end{abstract}

\begin{keyword}
neutron stars, equation of state, oscillations


\end{keyword}

\end{frontmatter}


\section{Introduction}
\label{sec:I}

Neutron stars, which are stellar remnants of the core collapse of
massive stars, give us one of the best opportunities to investigate
the physics under extreme conditions.  This is partly because the density 
of matter inside the star can become significantly larger than 
normal nuclear density, partly because some neutron stars are
strongly magnetized while others are rapidly rotating, and
partly because the gravitational fields around neutron stars are
strong enough for effects of general relativity to manifest 
themselves.  Astronomical phenomena associated with neutron stars could 
leave imprint of the physics under such extreme conditions.  The 
asteroseismology is a powerful technique to reveal the interiors of 
neutron stars via their oscillation spectra, as in the case of the 
seismology in the Earth and the helioseismology in the Sun.  In fact, 
through detection of the frequencies of neutron star oscillations, it 
might be possible to deduce neutron star masses and radii, the equation 
of state (EOS) of matter in the star, and so on (e.g., \cite{AK1996,STM2001,SKH2004,SYMT2011,DGKK2013}).  
Gravitational waves, if detected, could be one of the most promising 
sources that provide the oscillation spectra of neutron stars.

     Alternatively, it is expected that observational evidences for 
neutron star oscillations have already been given as the quasi periodic 
oscillations (QPOs) observed from soft gamma repeaters (SGRs) in the 
afterglow of the giant flares.  Up to now, at least three giant flares 
were observed, and various QPO frequencies were found in two
of them, which radiated from SGR 1806$-$20 and SGR 1900$+$14 
\cite{I2005,SW2005,SW2006}.   Since SGRs are considered to be 
magnetars, which are strongly magnetized neutron stars, the 
observed QPOs are expected to be strongly associated with the neutron 
star oscillations.  There are many attempts to explain these QPOs 
theoretically, which are based on shear torsional oscillations in the 
crustal region of a neutron star and/or magnetic oscillations 
throughout the star (e.g., \cite{Levin2006,Lee2007,SA2007,Sotani2007,Sotani2008a,Sotani2008b,Sotani2009}).  
Subsequently, analyses of elastic-magnetic oscillations in magnetars reveal 
that, depending on the magnetic field strength, the oscillations near the 
star's surface are excited basically by either crustal torsional oscillations or 
magnetic oscillations \cite{CK2011,GCFMS2012}.  According to observational 
estimates of the surface magnetic field strength of the SGRs from which 
the giant flares radiate \cite{K1998,H1999}, the magnetic field in the 
star can give rise to a restoring force, whose strength is comparable
to that induced by the crustal shear modulus, and hence shear torsional oscillations 
and magnetic oscillations are almost indistinguishable.  However, 
given that the highest magnetic fields are more or less localized, 
e.g., in a toroidal form,  it is reasonable to start with the assumption
that the QPOs observed in SGRs are identified as the crustal torsional oscillations.  
Then, such an identification could tell us information about the 
properties of the crust, particularly the EOS  \cite{SW2009,GNJL2011,S2011}.

     One of the most important parameters characterizing the EOS of 
neutron-rich matter in the crust is the density dependence of the nuclear 
symmetry energy, $L$, which is strongly associated with the thickness of the 
region where nuclei exist as non-uniform nuclear structures \cite{LRP1993,O1993,OI2007}.  
The constraint on $L$ can be given via the terrestrial nuclear
experiments \cite{Tsang2012}, but it is still difficult to obtain a 
severe constraint. On the other hand, we developed a way of constraining
$L$ through the identification of the QPOs observed in SGRs as the 
crustal torsional oscillations \cite{SNIO2012,SNIO2013a,SNIO2013b} as 
well as through possible simultaneous mass and radius measurements of low-mass
neutron stars \cite{SIOO2014}.  This is a constraint from nuclear 
matter at extremely large neutron excess, in contrast with the case of 
the terrestrial nuclear experiments associated with nuclear matter
at relatively small neutron excess.

     Recently, new QPO frequencies in SGR J1550$-$5418 and SGR 
1806$-$20 have been reported.  In SGR J1550$-$5418, the QPO frequencies of 
93 and 127 Hz have been discovered from a storm of 286 bursts 
\cite{QPO1}.  In addition, the QPO frequency of 57 Hz has been 
discovered in the shorter and less energetic recurrent 30 bursts radiating from
SGR 1806$-$20 \cite{QPO2}.  Since the QPO of frequency 57 Hz comes 
from the same SGR as that from which the giant flare was detected,
it is important to make sure that this new QPO frequency
can be explained within the 
framework that reproduces the low-lying QPO frequencies found in 
the giant flare. Meanwhile, although SGR J1550$-$5418 adds to a list of
the SGRs that have the QPOs detected, unfortunately the observed 
QPO frequencies are limited and not low enough to clearly identify
the corresponding angular indices.  Thus, we here systematically examine
how valid the crustal torsional oscillations are to explain the QPO frequencies
observed in the SGRs.  In order to calculate the frequencies of the
crustal torsional oscillations, we adopt the shear modulus that allows
for the effect of electron screening on nuclei in the crust
\cite{KP2013}.  Additionally, as the effect of neutron superfluidity 
on the enthalpy density of matter in the inner crust, we adopt the 
results for the superfluid density given by Chamel \cite{Chamel2012}, as in Refs.
\cite{SNIO2013a,SNIO2013b}.  We adopt the geometric unit of $c=G=1$ in this 
paper, where $c$ and $G$ denote the speed of light and the gravitational 
constant, respectively.

\section{Crust configuration and EOS parameters}
\label{sec:II}

     In the vicinity of the saturation point of symmetric nuclear matter 
at zero temperature, the bulk energy of nuclear matter per nucleon $w$ 
can be generally expressed as a function of baryon number density $n_{\rm b}$ 
and neutron excess $\alpha$, as in Ref. \cite{L1981}:
\begin{equation}
  w = w_0  + \frac{K_0}{18n_0^2}(n_{\rm b}-n_0)^2 + \left[S_0 
      + \frac{L}{3n_0}(n_{\rm b}-n_0)\right]\alpha^2,
  \label{eq:w}
\end{equation}
where $w_0$, $n_0$, and $K_0$ are the saturation energy, saturation density, 
and incompressibility of symmetric nuclear matter ($\alpha=0$). In addition, 
$S_0$ and $L$ are associated with the density dependent symmetry energy 
$S(n_{\rm b})$ as $S_0\equiv S(n_0)$ and $L\equiv 
3n_0(dS/dn_{\rm b})_{n_{\rm b}=n_0}$.  Note that only $w_0$, $n_0$, and $S_0$ 
among the five parameters are well constrained from empirical 
data for masses and radii of stable nuclei \cite{OI2003}.

    In describing matter in the crust of a neutron star, we follow the derivation
by two of the authors (K.O. and K.I.),
which is based on a phenomenological approach \cite{OI2007}.
Hereafter, we refer to the resultant phenomenological EOS of matter in the crust as 
OI-EOS.  The OI-EOS is constructed as follows.  First, various 
models for the bulk energy $w(n_{\rm b},\alpha)$ of nuclear matter are
made in such a way as to reproduce Eq.\ (\ref{eq:w}) in the limit of 
$n_{\rm b}\to 0$ and $\alpha\to 0$.  Then, within a simplified version 
of the extended Thomas-Fermi theory, the density profile of stable nuclei 
was obtained for each model for $w$.  Finally, the optimal values of 
$w_0$, $n_0$, and $S_0$ were determined so that the charge number, mass 
excess, and charge radius that can be calculated from the density 
profile obtained for given $y\equiv -K_0S_0/(3n_0L)$ and $K_0$ 
should fit well to the experimental data \cite{OI2003}.  After that, 
in order to obtain the equilibrium nuclear shape and size as well as the 
crust EOS for various sets of $y$ and $K_0$, we generalized the 
Thomas-Fermi model by adding dripped neutrons, a neutralizing  
uniform background of electrons, and the lattice energy within a 
Wigner-Seitz approximation \cite{OI2007}.  In the present work, as in Refs. 
\cite{OI2007,SNIO2012,SNIO2013a,SNIO2013b}, we consider the parameters 
$L$, $K_0$, and $y$ in the range of $0<L<160$ MeV, $180\le K_0\le 360$ MeV, 
and $y<-200$ MeV fm$^3$, which can not only reproduce the mass and radius 
data for stable nuclei equally well, but also effectively cover even 
extreme cases \cite{OI2003}.  The eleven parameter sets adopted in 
this work are the same as in Table 1 in Ref. \cite{SNIO2013b}.

     For given mass $M$ and radius $R$ of a nonrotating neutron star,
the crust configuration can be constructed by integrating the 
Tolman-Oppenheimer-Volkoff equations with the above-mentioned crust EOS
from the star's surface inward up to the basis of the crust 
\cite{IS1997,SNIO2012,SNIO2013a,SNIO2013b}.  This construction 
effectively avoids uncertainties in the core EOS.  In this work, we 
particularly consider typical neutron star models with $M$ and $R$ 
in the range of $1.4\le M/M_\odot\le 1.8$ and $10$ km $\le R \le 14$ km.

     One of the crucial properties that govern the shear 
oscillations is the elasticity, which is characterized by the shear modulus 
$\mu$.  The shear modulus in the crust is mainly determined by the lattice 
energy due to the Coulomb interaction, which is approximately given as
\begin{equation}
  \mu = 0.1194\times\frac{n_i(Ze)^2}{a}, \label{eq:off}
\end{equation}
where $n_i$, $Z$, and $a=(3/4\pi n_i)^{1/3}$ are the ion number density,  
the nuclear charge number, and the radius of a Wigner-Seitz cell, 
respectively \cite{SHOII1991}.  This formula is derived in the limit of zero 
temperature from Monte Carlo calculations averaged over all directions on the 
assumption that the nuclei are point particles forming the body-center cubic 
lattice \cite{OI1990}.

\begin{figure}
\begin{center}
\includegraphics[scale=0.5]{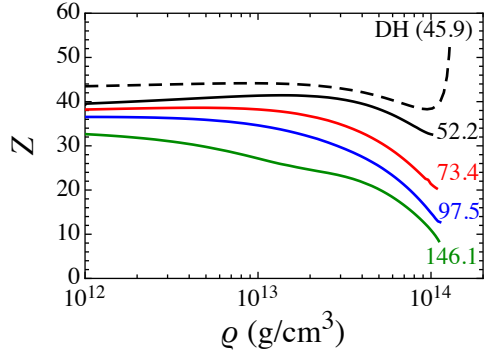} 
\end{center}
\caption{
(Color online) Nuclear charge number, $Z$, plotted as a function 
of the energy density, $\rho$, for the OI-EOS with various values of $L$.  The 
labels on the solid lines denote the values of $L$; $L=52.2$, 73.4, 97.5, and 
146.1 MeV correspond to the OI-EOS with $(y, K_0)$=($-220$ MeV fm$^3$, 
180 MeV), ($-220$ MeV fm$^3$, 230 MeV), ($-220$ MeV fm$^3$, 280 MeV), and 
($-220$ MeV fm$^3$, 360 MeV).  For reference, the result from the EOS
proposed by Douchin \& Haensel \cite{DH} is also shown with the broken line, 
where the corresponding value of $L$ is 45.9 MeV.
}
\label{fig:Z}
\end{figure}

    On the other hand, Kobyakov and Pethick proposed the modification of the shear 
modulus, where the effect of electron screening is taken into account \cite{KP2013}.
The modified shear modulus can be expressed as
\begin{equation}
  \mu = 0.1194\left(1-0.010Z^{2/3}\right)\frac{n_i(Ze)^2}{a}, \label{eq:on}
\end{equation}
where the term of $Z^{2/3}$ represents the effect of electron screening. 
That is, the shear modulus decreases due to such additional effect, 
which in turn leads to reduction of the frequencies of torsional 
oscillations \cite{S2014}\footnote{The electron screening acts to change
the toroidal oscillation frequencies also via modifications of the enthalpy
density and the crustal structure, but such change is negligibly small.}.  
Since the charge number $Z$ also depends on $L$, 
however, it is still unclear how the reduction of the frequencies of 
torsional oscillations due to the effect of electron screening depends on 
$L$.  At subnuclear densities, $Z$ tends to decrease with $L$, because 
the smaller symmetry energy, corresponding to larger $L$, helps 
more protons to change into neutrons.  In fact, $Z$ decreases with 
$L$ for the OI-EOS adopted in the present work, as shown in Fig.\ 
\ref{fig:Z}.  Consequently, one expects that the frequencies of torsional 
oscillations by using the modified shear modulus [Eq. (\ref{eq:on})]  
differs little from those by using the original shear modulus 
[Eq. (\ref{eq:off})], if $L$ is sufficiently large.  Even so, we will 
examine the torsional oscillations with the modified shear modulus 
[Eq. (\ref{eq:on})] in this work.

\section{Crustal torsional oscillations in neutron stars}
\label{sec:III}

     We proceed to calculate the fundamental frequencies of the 
torsional oscillations that are excited in the crust of a nonrotating
neutron star of mass $M$ and radius $R$ as constructed from the crust 
EOS of a given set of $L$ and $K_0$ in the previous section.  In particular, 
we adopt the relativistic Cowling approximation to calculate the 
frequencies.  That is, we neglect the metric perturbations and keep them 
zero during the oscillations.  This is presumably a good approximation 
for considering the torsional oscillations, because they are axial 
parity oscillations, which do not involve the density perturbations.  The 
perturbation equation for the torsional oscillations can be obtained 
from the linearized relativistic equation of motion \cite{ST1983}.  Then, 
imposing the appropriate boundary conditions at the top and bottom of
the crust, the problem to solve reduces to the eigenvalue problem, where the 
eigenvalues correspond to the eigen-frequencies of the torsional 
oscillations. The perturbation equation and the boundary conditions are 
explicitly described in Ref. \cite{SNIO2013b}; pasta nuclei, if present,
are assumed to have zero shear modulus.

     Additionally, the effect of dripped neutrons should be taken into account 
for the calculation of the frequencies of the torsional oscillations. 
In fact, neutrons are generally considered to drip out of nuclei when 
the density becomes higher than $\sim 4\times 10^{11}$ g/cm$^3$, and some of 
them behave as a superfluid.  Unfortunately, it is still unclear 
how many dripped neutrons behave as such, but most of them 
can move together with the nuclei because of the entrainment 
effects \cite{Chamel2012}.  In fact, according to the results from the 
band calculations in Ref. \cite{Chamel2012}, only of the order of $10-30$ \% of the 
dripped neutrons can participate in superfluidity at 
$n_{\rm b} \sim 0.01-0.4n_0$.  Through the enthalpy density, the 
frequencies of the torsional oscillations depends strongly on this ratio 
\cite{SNIO2013a}; we adopt the results by Chamel \cite{Chamel2012} in the 
present work.  Then, we calculate the frequencies based on the prescription 
how to build the effect of superfluidity into the perturbation equations 
shown in Ref. \cite{SNIO2013b}.

\begin{figure}
\begin{center}
\includegraphics[scale=0.5]{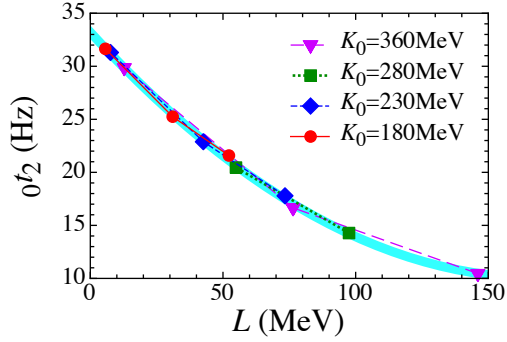} 
\end{center}
\caption{
(Color online) The frequency of the fundamental torsional oscillations 
with $\ell=2$, ${}_0t_2$, plotted as a function of $L$ for the stellar models 
with $M=1.4M_\odot$ and $R=12$ km.  The thick solid line denotes the fitting 
formula [Eq. (\ref{eq:fit2})]. 
}
\label{fig:0t2}
\end{figure}

     First, to see the dependence of the frequencies on the parameters 
that characterize the crust EOS, $L$ and $K_0$, we calculate the frequencies 
of the fundamental torsional oscillations with $\ell=2$ for the stellar models 
with $M=1.4M_\odot$ and $R=12$ km.  The calculated frequencies are plotted as a
function of $L$ for various values of $K_0$ in Fig.\ \ref{fig:0t2}. 
Note that the shear torsional oscillations are often referred to as
$t$-modes, which are labelled as ${}_nt_\ell$ with the angular index $\ell$ and
the number $n$ of radial nodes in the eigenfunction.  From this figure, 
one can observe that the frequency ${}_0t_2$ is almost independent of 
the incompressibility $K_0$, while depending strongly on $L$. This 
tendency has been already shown in the case in which ${}_nt_\ell$ is
calculated from the shear modulus [Eq. (\ref{eq:off})] in Ref. \cite{SNIO2013a}, but 
still holds in the case of the modified shear modulus [Eq. (\ref{eq:on})]. 
Since we confirm that this is true for different stellar models within 
$1.4\le M/M_\odot \le 1.8$ and $10$ km $\le R\le 14$ km, ${}_0t_2$ can be 
approximately expressed as
\begin{equation}
  {}_0t_2 = c_2^{(0)} - c_2^{(1)}L + c_2^{(2)}L^2, \label{eq:fit2}
\end{equation}
where $c_2^{(0)}$, $c_2^{(1)}$, and $c_2^{(2)}$ denote positive coefficients  
that depend on the stellar models.  In practice, this fitting formula agrees 
well with the calculated frequencies for various sets of the EOS 
parameters within the accuracy of a few per cent, as shown in Table 
\ref{tab:0t2-M14R12}.

\begin{table*}
\centering
\caption{
The numerically calculated values of the frequency of the $\ell=2$ 
fundamental torsional oscillations, ${}_0t_2^{(c)}$, and the values 
obtained from Eq.\ (\ref{eq:fit2}), ${}_0t_2^{(e)}$, for the stellar 
models of $M=1.4M_\odot$ and $R=12$ km.  The relative errors determined by 
$({}_0t_2^{(c)}-{}_0t_2^{(e)})/{}_0t_2^{(c)}$ are also tabulated.
}
\footnotesize
\begin{tabular}{cccccc}
\hline\hline
 $y$ (MeV fm$^3$) & $K_0$ (MeV) & $L$ (MeV) & ${}_0t_2^{(c)}$ (Hz) & ${}_0t_2^{(e)}$ (Hz) & relative error (\%) \\
\hline
  $-220$ & 180 & 52.2 & 21.59 & 21.12 & 2.14  \\
  $-220$ & 230 & 73.4 & 17.79 & 17.49 & 1.71  \\
  $-220$ & 280 & 97.5 & 14.28 & 14.21 & 0.43  \\
  $-220$ & 360 & 146.1 & 10.46 & 10.47 & $-0.15$  \\
  $-350$ & 180 & 31.0 & 25.25 & 25.49 & $-0.98$  \\
  $-350$ & 230 & 42.6 & 22.89 & 23.02 & $-0.60$  \\
  $-350$ & 280 & 54.9 & 20.45 & 20.63 & $-0.86$  \\
  $ -350$ & 360 & 76.4 & 16.65 & 17.02 & $-2.21$  \\
  $-1800$ & 180 & 5.7 & 31.65 & 31.68 & $-0.12$  \\
  $-1800$ & 230 & 7.6 & 31.30 & 31.17 & 0.41  \\
  $-1800$ & 360 & 12.8 & 29.87 & 29.84 & 0.08  \\
\hline\hline
\end{tabular}
\label{tab:0t2-M14R12}
\end{table*}

     In a similar way, we find that the frequencies ${}_0t_\ell$ for 
$\ell>2$ are almost independent of $K_0$ and can be approximately expressed 
as a function of $L$ via
\begin{equation}
  {}_0t_\ell = c_\ell^{(0)} - c_\ell^{(1)}L + c_\ell^{(2)}L^2, \label{eq:fitl}
\end{equation}
where $c_\ell^{(0)}$, $c_\ell^{(1)}$, and $c_\ell^{(2)}$ are adjustable 
parameters.

\begin{figure}
\begin{center}
\includegraphics[scale=0.5]{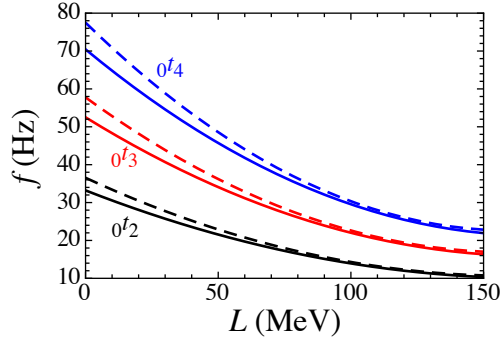} 
\end{center}
\caption{
(Color online) 
The frequencies of the fundamental torsional oscillations with $\ell=2$, 3, 
and 4, plotted as a function of $L$ for the stellar models with 
$M=1.4M_\odot$ and $R=12$ km.  The broken lines correspond to the frequencies 
calculated from the shear modulus [Eq. (\ref{eq:off})], while the solid lines 
correspond to those from Eq. (\ref{eq:on}).
}
\label{fig:screening}
\end{figure}

     In Fig.\ \ref{fig:screening}, we show the comparison between the 
frequencies calculated from the shear modulus [Eq. (\ref{eq:off})] 
and those from the modified shear modulus [Eq. (\ref{eq:on})] for 
the stellar models with $M=1.4M_\odot$ and $R=12$ km.  As 
discussed in the previous section, the deviations in 
${}_0t_\ell$ due to the effect of electron screening increase
with decreasing $L$, but almost vanish for large $L$.

\section{Comparison with the QPO frequencies}
\label{sec:IV}

\begin{figure}
\begin{center}
\includegraphics[scale=0.5]{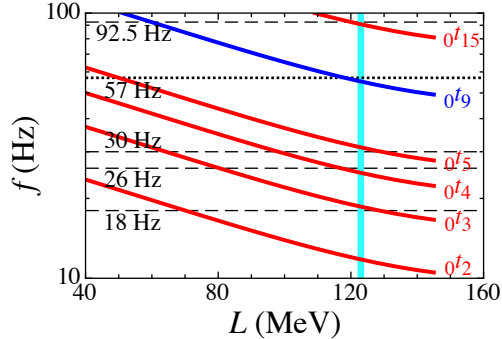} 
\end{center}
\caption{
(Color online) Correspondence between the QPO frequencies observed in SGR 
1806$-$20 and the calculated frequencies of the fundamental torsional 
oscillations for the stellar models with $M=1.4M_\odot$ and $R=12$ km.  
The horizontal broken lines denote the QPO frequencies originally
reported by \citet{I2005,SW2006}, while the horizontal dotted line denotes 
the new QPO frequency discovered by \citet{QPO2}.  The vertical line 
denotes the value of $L=123$ MeV, with which the observed QPO frequencies 
agree best with the calculated frequencies of the crustal torsional oscillations.  
}
\label{fig:1806-M14R12}
\end{figure}

     Let us now compare the calculated frequencies ${}_0t_\ell$ with the QPO 
frequencies observed in SGRs.  The identification of the QPO frequencies observed 
in SGR 1806$-$20 as the frequencies ${}_0t_\ell$ is more difficult than that 
in SGR 1900$+$14, because not only many QPO frequencies are discovered in SGR 
1806$-$20 but also the interval between the observed QPO frequencies 26 and 30 Hz 
is remarkably small \cite{Sotani2007}.  Nevertheless, as shown in Refs. 
\cite{SNIO2013a,SNIO2013b}, the QPO frequencies 18, 26, 30, and 92.5 Hz observed 
in SGR 1806$-$20 can be explained well in terms of the fundamental torsional 
oscillations with $\ell=3, 4, 5$, and 15, where the frequencies of the torsional 
oscillations are calculated from the original shear modulus [Eq. (\ref{eq:off})]. 
Here, we re-calculate the frequencies from  the modified shear modulus 
[Eq. (\ref{eq:on})].  Then, we find that the same correspondence as in  Refs. 
\cite{SNIO2013a,SNIO2013b} is still intact, as shown in Fig.\ 
\ref{fig:1806-M14R12} for the stellar models with $M=1.4M_\odot$ and $R=12$ 
km.  We remark that the reproduction of the QPO frequencies by the crustal 
torsional oscillations is equally good.  Moreover, we also include
the new QPO frequency 57 Hz reported by \cite{QPO2} in this figure, and we find 
that this additional QPO frequency can be identified as the $\ell=9$ 
fundamental torsional oscillations.  As can be seen from Fig.\ 
\ref{fig:1806-M14R12}, the most suitable value of $L$ to adjust the crustal 
torsional oscillations to the five observed QPO frequencies is $L=123$ MeV, 
with which the frequencies obtained from Eq.\ (\ref{eq:fitl}) and their 
relative errors from the observed QPO frequencies are shown in Table 
\ref{tab:1806-M14R12}.  Since not only the QPO frequencies discovered from the 
giant flare but also that observed from the different event in the same object, 
SGR 1806$-$20, can be explained in the same framework, we conclude that 
the torsional oscillations are still promising as the origin of the
QPOs from SGRs.

\begin{table}
\centering
\caption{The QPO frequencies observed in SGR 1806$-$20, the corresponding 
angular indices and frequencies of the fundamental torsional oscillations, 
the latter of which are obtained from Eq.\ (\ref{eq:fitl}) 
by substituting the optimal value into $L$ for
$M=1.4M_\odot$, and $R=12$ km, and their relative errors
from the observed QPO frequencies.}
\footnotesize
\begin{tabular}{cccc}
\hline\hline
 QPO frequency (Hz) & $\ell$ & ${}_0t_\ell^{(e)}$ (Hz) & relative error (\%) \\
\hline
   $18$ & 3 & 18.62 & $-3.44$   \\
   $26$ & 4 & 24.98 & $3.92$    \\
   $30$ & 5 & 31.15 & $-3.85$    \\
   $57$ & 9 & 55.22 & $3.13$    \\
   $92.5$ & 15 & 90.76 & $1.88$    \\
\hline\hline
\end{tabular}
\label{tab:1806-M14R12}
\end{table}

\begin{figure}
\begin{center}
\includegraphics[scale=0.5]{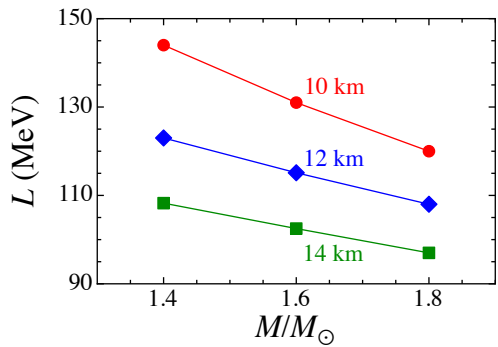} 
\end{center}
\caption{
(Color online) The optimal values of $L$ to explain the QPO 
frequencies observed in SGR 1806$-$20, as in Fig.\ \ref{fig:1806-M14R12}, 
for various stellar models.  In the figure, the circles, diamonds, and 
squares correspond to the results for the stellar models with $R=10$, 12, 
and 14 km. 
}
\label{fig:L-1806}
\end{figure}

     The same identification between the QPO frequencies observed in SGR 
1806$-$20 and the crustal torsional oscillations is also possible for 
different stellar models.  In Fig.\ \ref{fig:L-1806}, we plot the 
optimal values of $L$ to explain the observed QPO frequencies for various 
stellar models within $1.4\le M/M_\odot \le 1.8$ and 10 km $\le R \le$ 14 km.

\begin{figure}
\begin{center}
\includegraphics[scale=0.5]{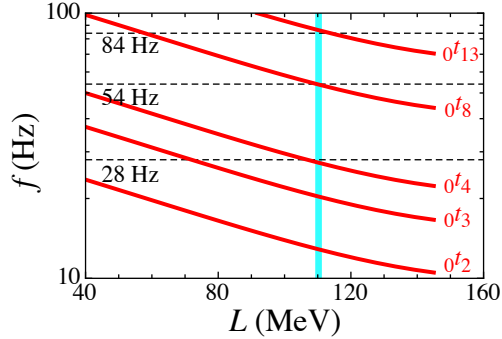} 
\end{center}
\caption{
(Color online) Same as Fig.\ \ref{fig:1806-M14R12}, but for the QPO 
frequencies observed in SGR 1900$+$14.
}
\label{fig:1900-M14R12}
\end{figure}

\begin{figure}
\begin{center}
\includegraphics[scale=0.5]{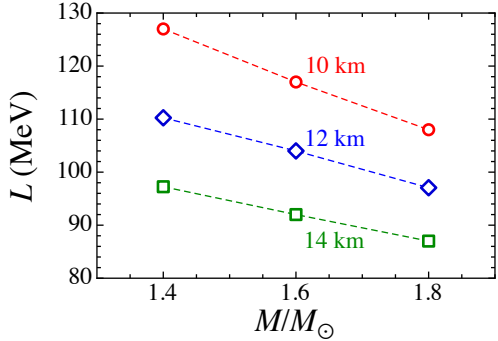} 
\end{center}
\caption{
(Color online) Same as Fig.\ \ref{fig:L-1806}, but for the QPO 
frequencies observed in SGR 1900$+$14.
}
\label{fig:L-1900}
\end{figure}

     Unlike SGR 1806$-$20, the observed QPO frequencies in SGR 1900$+$14 
can be simply explained in terms of the crustal torsional 
oscillations.  As in Ref. \cite{SNIO2013a}, we find that the observed  
QPOs of frequencies 28, 54, and 84 Hz can be identified as 
$\ell=4, 8$, and 13 even for the calculations from the modified 
shear modulus (\ref{eq:on}).  In Fig.\ \ref{fig:1900-M14R12}, we show 
such an identification for the stellar models with $M=1.4M_\odot$ 
and $R=12$ km, together with the optimal value of $L$ to explain 
the observed QPO frequencies, i.e., $L=110$ MeV.  We also show the 
comparison between the observed QPO and calculated frequencies for the 
stellar models with $M=1.4M_\odot$ and $R=12$ km in Table 
\ref{tab:1900-M14R12}.  Additionally, we find that the observed QPO 
frequencies can be explained by the same identification even for various 
stellar models, where the corresponding optimal value of $L$ 
to explain the observed QPO frequencies in SGR 1900$+$14 is
shifted as shown in Fig.\ \ref{fig:L-1900}.

\begin{table}
\centering
\caption{Same as Table \ref{tab:1806-M14R12}, but for the QPO 
frequencies observed in SGR 1900$+$14.}
\footnotesize
\begin{tabular}{cccc}
\hline\hline
 QPO frequency (Hz) & $\ell$ & ${}_0t_\ell^{(e)}$ (Hz) & relative error (\%) \\
\hline
   $28$ & 4 & 27.29 & $2.55$   \\
   $54$ & 8 & 53.80 & $0.36$    \\
   $84$ & 13 & 86.26 & $-2.69$    \\
\hline\hline
\end{tabular}
\label{tab:1900-M14R12}
\end{table}

\begin{figure}
\begin{center}
\includegraphics[scale=0.5]{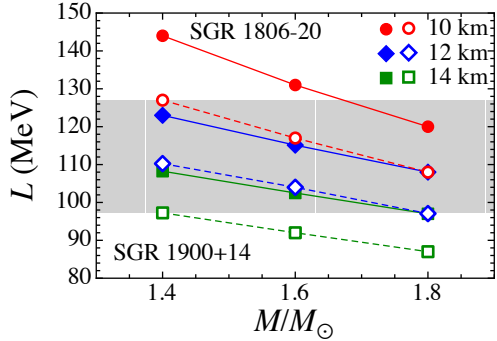} 
\end{center}
\caption{
(Color online) The allowed values of $L$ (shaded region) that can 
simultaneously explain the QPO frequencies observed in SGR 1806$-$20 and 
1900$+$14, if the corresponding central objects are neutron stars  
having mass and radius in the range of $1.4\le M/M_\odot \le 1.8$ and 
$10$ km $\le R\le 14$ km.
}
\label{fig:LL}
\end{figure}

     Although the value of $L$ still depends on many 
uncertainties, it is reasonable to try to simultaneously explain 
the QPO frequencies observed in SGR 1806$-$20 and SGR 1900$+$14 
for a specific value of $L$.  As shown in Fig.\ \ref{fig:LL}, 
therefore, we can constrain $L$ as $97\le L\le 127$ MeV 
by assuming that the central objects in SGR 1806$-$20 and 
SGR 1900$+$14 are neutron stars whose mass and radius  
are in the range of $1.4\le M/M_\odot \le 1.8$ and $10$ km 
$\le R\le 14$ km.  Comparing to the constraint on $L$ obtained in 
Ref. \cite{SNIO2013b} from the shear modulus [Eq. (\ref{eq:off})], the 
current constraint shifts to lower values by about 4 MeV
due to the effect of electron screening.  We also remark that the 
obtained constraint is still consistent with that from an X-ray 
busting neutron star \cite{SIO2015}.

     Furthermore, we proceed to the QPO frequencies observed in SGR 
J1550$-$5418, which give us information independent of SGR 1806$-$20 and 
SGR 1900$+$14.  Unfortunately, there are only two QPO frequencies 
observed, 93 and 127 Hz \cite{QPO1}.  The observed frequencies, which
are relatively high, might arise from the same oscillation mechanism 
other than the crustal torsional oscillations, but, as we shall 
see, the crustal torsional oscillations have little difficulty in 
reproducing the observed QPO frequencies.

\begin{figure}
\begin{center}
\includegraphics[scale=0.5]{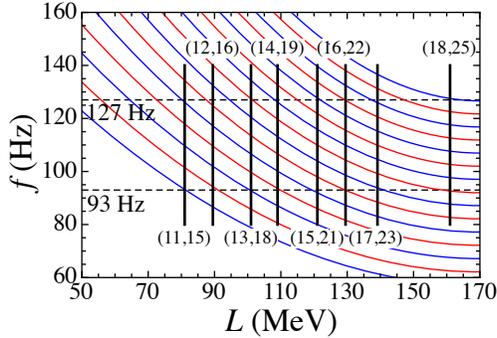} 
\end{center}
\caption{
(Color online) The frequencies of the fundamental torsional oscillations with 
$\ell=11-25$, which are evaluated as a function of $L$ for the 
stellar models with $M=1.4M_\odot$ and $R=12$ km.  The horizontal broken 
lines denote the QPO frequencies observed in SGR J1550$-$5418.  The vertical 
lines denote various possible correspondences between the QPO frequencies
and the frequencies of the fundamental torsional oscillations with the 
values of $\ell$ represented by $(*,*)$ in the figure.
}
\label{fig:1550-M14R12}
\end{figure}

     In Fig.\ \ref{fig:1550-M14R12}, the frequencies of the fundamental 
torsional oscillations with $\ell=11-25$ are shown as a function of $L$ for 
the stellar models with $M=1.4M_\odot$ and $R=12$ km, together with the 
observed QPO frequencies.  From this figure, one finds various 
combinations of ${}_0t_\ell$ that can reproduce the observed QPO 
frequencies.  In practice, if the observed QPO frequencies 93 Hz and 127 Hz are
regarded as the frequencies of the fundamental torsional oscillations 
with $\ell_1$ and $\ell_2 (>\ell_1)$, one can find that $(\ell_1, \ell_2)=(11, 15), (12, 16), (13, 18), (14, 19), (15, 21), (16, 22), (17, 23)$, and 
$(18, 25)$ are relevant as shown by the vertical solid lines in 
Fig.\ \ref{fig:1550-M14R12}, where the corresponding optimal values of $L$
are $81$, $90$, $101$, $109$, $121$, $130$, $139$, and $161$ MeV, respectively.
Additionally, in Table \ref{tab:1550-M14R12}, we 
show the possible combinations of $\ell_1$ and $\ell_2$, 
the frequencies of the corresponding torsional oscillations
that can be obtained from Eq. (\ref{eq:fitl}) by substituting the optimal
values into $L$ for $M=1.4M_\odot$ and $R=12$ km, and their relative errors 
from the observed QPO frequencies.  Considering the accuracy of the 
fitting formula (\ref{eq:fitl}) as well as the limited number of the 
observed QPO frequencies, we have difficulty in judging which 
combination is in best agreement with the observed QPO frequencies.

\begin{table}
\centering
\caption{Same as Table \ref{tab:1806-M14R12}, but for the QPO frequencies 
observed in SGR J1550$-$5418, where various combinations of $\ell$ to 
explain the observed QPO frequencies can be considered.
}
\footnotesize
\begin{tabular}{cccc}
\hline\hline
 QPO frequency (Hz) & $\ell$ & ${}_0t_\ell^{(e)}$ (Hz) & relative error (\%) \\
\hline
   $93$ & 11 & 93.19 & $-0.21$   \\
   $127$ & 15 & 126.1 & $0.73$    \\ \hline
   $93$ & 12 & 94.26 & $-1.35$   \\
   $127$ & 16 & 124.8 & $1.74$    \\ \hline
   $93$ & 13 & 92.66 & $0.37$   \\
   $127$ & 18 & 127.4 & $-0.28$    \\ \hline
   $93$ & 14 & 93.59 & $-0.63$   \\
   $127$ & 19 & 126.1 & $0.70$    \\ \hline
   $93$ & 15 & 91.94 & $1.14$   \\
   $127$ & 21 & 127.7 & $-0.57$    \\ \hline
   $93$ & 16 & 92.91 & $0.098$   \\
   $127$ & 22 & 126.8 & $0.15$    \\ \hline
   $93$ & 17 & 93.77 & $-0.83$   \\
   $127$ & 23 & 125.9 & $0.84$    \\ \hline
   $93$ & 18 & 92.61 & $0.42$   \\
   $127$ & 25 & 127.5 & $-0.40$    \\
\hline\hline
\end{tabular}
\label{tab:1550-M14R12}
\end{table}

\begin{figure}
\begin{center}
\includegraphics[scale=0.5]{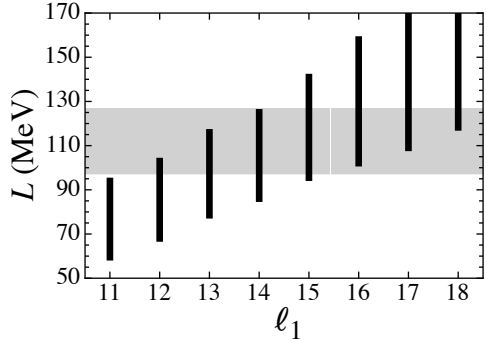} 
\end{center}
\caption{
The allowed regions of $L$ to explain the QPO frequencies 
observed in SGR J1550$-$5418 in terms of the fundamental torsional 
oscillations for the stellar models with the mass and radius in the range of 
$1.4\le M/M_\odot \le 1.8$ and $10$ km $\le R\le 14$ km.  The horizontal axis 
$\ell_1$ denotes the angular index as which the lower QPO frequency 93 
Hz is identified, as shown in Table \ref{tab:1550-M14R12}.  The shaded region 
is the constraint on $L$ from the QPO frequencies observed in SGR 
1806$-$20 and SGR 1990$+$14 as described in Fig.\ \ref{fig:LL}.
We set the upper limit of the vertical axis at $L=170$ MeV,
above which extrapolations based on Eq.\ (\ref{eq:fitl}) are no longer
effective.
}
\label{fig:LL3}
\end{figure}

     Assuming that the central object in SGR J1550$-$5418 is a neutron star 
within $1.4\le M/M_\odot \le 1.8$ and $10$ km $\le R\le 14$ km, we show in 
Fig.\ \ref{fig:LL3} that several combinations of $(\ell_1,\ell_2)$ are 
consistent with the constraint on $L$ from the QPO frequencies observed in
SGR 1806$-$20 and SGR 1900$+$14 as described in Fig.\ \ref{fig:LL}.  As 
in the case of SGR 1806$-$20 and SGR 1900$+$14 shown in Figs.\ 
\ref{fig:L-1806} and \ref{fig:L-1900}, the upper and lower limits of the 
constraint on $L$ for each $(\ell_1,\ell_2)$ in Fig.\ \ref{fig:LL3} are 
determined by the stellar models with $(M, R)$=(1.4$M_\odot$, 10 km) and 
(1.8$M_\odot$, 14 km), respectively.  From this figure, we find that at 
least the identification of the QPOs observed in SGR J1550$-$5418 
as the fundamental torsional oscillations with $\ell =11$ and 15 is not 
consistent with the constrained region of $L$ from the QPO frequencies
observed in SGR 1806$-$20 and SGR 1900$+$14.  Additional discoveries of the 
QPO frequencies in SGR J1550$-$5418 would enable us to tell which 
of the remaining possibilities is best, which in turn leads to 
severer constraint on $L$.

\section{Conclusion}
\label{sec:V}

   In summary, we show that the observed low-lying QPOs, including
the new ones from SGR 1806$-$20 and SGR J1550$-$5418, can be still 
identified as the crustal fundamental torsional oscillations with different 
$\ell$'s.  This result gives a strong indication that the QPOs and the 
crustal modes are more or less related.  However, there remain many problems.
To make better estimates of the frequencies of the crustal modes, which are
given as a function of the EOS parameter $L$ and the neutron star mass $M$
and radius $R$ in the present work, it is indispensable to take into account 
effects of magnetic fields on the effective shear modulus.  Unfortunately, it 
would be a tall task given poorly known magnetic structure.  Even within the 
framework of purely elastic shear modes as considered in the present work, 
how to average the locally anisotropic shear modulus over directions for a 
polycrystal could modify the effective shear modulus significantly 
\cite{KP2015}.  Also, stability of the bcc structure as assumed here is 
endangered by fluctuations in the density of dripped neutrons; the resultant 
change in the lattice structure could affect the effective shear modulus and 
modifications of the enthalpy density by neutron superfluidity \cite{KP2014}.
Note that with all those theoretical uncertainties, one can definitely assign 
the angular index $\ell$ to each QPO except the two QPOs observed in SGR 
J1550$-$5418; assignment of these two QPOs would be possible only after 
possible observations of additional low-lying QPOs from the same source.

This work was supported in part by Grants-in-Aid for Scientific Research on 
Innovative Areas through No.\ 15H00843 and 
No.\ 24105008 provided by MEXT and
by Grant-in-Aid for Young Scientists (B) through No.\ 26800133 provided by JSPS.









\end{document}